# Signal and Charge Transfer Efficiency of Few Electrons Clocked on Microscopic Superfluid Helium Channels

G. Sabouret, F. R. Bradbury, S. Shankar and S. A. Lyon

Department of Electrical Engineering, Princeton University, Princeton, New Jersey

08544

Electrons floating on the surface of liquid helium are possible spin-qubits for quantum information processing. Varying electric potentials are not expected to modify spin states, which allows their transport on helium using a charge-coupled device (CCD)-like array of underlying gates. This approach depends upon efficient inter-gate transfer of individual electrons. Measurements are presented here of the charge transfer efficiency (CTE) of few electrons clocked back and forth above a short microscopic CCD-like structure. A charge transfer efficiency of 0.99999992 is obtained for a clocking frequency of 800 kHz.

73.20.-r, 73.25.+i, 73.90.+f, 67.90.+z, 03.67.Lx

Electrons floating on the surface of superfluid helium are potential candidates as qubits for quantum computation[1,2] relying either on their charge states[3,4,5,6] or spin states[7] as two-level quantum systems. In the case of spin qubits, the electrons would be moved about the surface of the helium using a charge coupled device[8] (CCD)-like network of underlying gates, making this scheme intrinsically scalable. A well-known concern for silicon CCD's is their limited charge transfer efficiency (CTE) due to charge trapping at the Si/SiO$_2$ interface or in bulk traps[9]. Whereas leaving a few electrons behind on the clocking path is tolerable in silicon CCD's due to the large number of electrons involved, it becomes unacceptable for quantum computing where single electrons must be transported without loss. We have previously reported measurements of the charge transfer efficiency in a macroscopic CCD structure covered by a millimeter-thick layer of helium and found that the inter-gate transfer was governed by electron diffusion[10]. Here we report the measurement of the CTE for few electrons on microscopic superfluid helium channels in a short CCD. Using a clocking technique that expels electrons which do not transfer properly, we measure an exceptionally large CTE of 0.99999992.

The sample used in this experiment is a sapphire substrate on which a multilayer metal/polymer/metal structure is deposited. The lower metallic layer is patterned into the underlying gates using photolithography, while the top metal and polymer layers are cut into an array of sixty 10 µm-wide channels using photolithography followed by reactive ion and wet chemical etching. Figure 1(a) is a picture of the device seen from the top with (b) a schematic vertical cross-section showing where the helium and electrons lie.

The device is mounted in a sealed copper cell held at a temperature of 1.55K in a pumped $^4$He system. A mass flowmeter is used to determine the quantity of $^4$He gas admitted and sets the level of bulk liquid helium inside the cell to a few millimeters below the sample. The channels are filled by capillary action and the helium thickness is therefore set accurately by the height of the ridges of the channels[11]. A schematic of the interior of the cell is shown in figure 2. A sapphire window coated with Ti/Zn is placed 4 mm above the device. The channel gates on the sample are maintained at direct current (dc) ground, while the device's top plane in which the channels are patterned is kept at –2 V. At the beginning of the experiment, a -3V potential is applied to The Ti/Zn and a pulse of UV light from the optical fiber photo-emits electrons onto the sample[12]. Besides the zero dc voltage bias common to all seven small gates, independent alternating current (ac) signals can be applied to gates 3 through 7. Electrons above gate 1 capacitively couple to it for detection. Gate 1 is connected to a lock-in amplifier through a HEMT preamp located inside the cell which acts as a capacitance buffer (we are dealing with very few electrons, so the ~400 pF capacitance of the wires running all the way up to room temperature would make their signal quite small). In addition to the seven small gates (seen in the inset of Fig. 1), there are three large gates (400 μm wide, separated from the small gates by a small "door gate ", D) that act as a reservoir where photoemitted electrons are collected. These large gates can also be used to verify the presence of electrons when connected in a Sommer-Tanner three-plate configuration[13].

We first ran a conventional three-phase CCD gate voltage sequence (alternating between driving the electrons to the left and right) on gates 3 through 7 to move electrons back and forth between gate 1 and gate 7. This is the same sequence used previously with mm-thick He and large area gates.[10] The potential on gate 2 is held constant so that it acts as an ac ground to help shield gate 1 from any signal coupled from the changing voltages on the other gates. The clocking sequence moves electrons towards gate 7 for three sub-periods, pauses for three sub-periods, moves the electrons towards gate 1 for three sub-periods, pauses again for three sub-periods and repeats this cycle. The pauses improve the detection sensitivity. The reference frequency of the lock-in amplifier is set to be equivalent to a round trip cycle of the electrons (12 sub-periods). The ac clocking voltages are 450mV, peak-to-peak. The large gates P1, P2 and P3 are maintained at 0V like the dc components of the small gates, and gate D is set at –2 V like the top plane to separate the large and small gates.

To begin the experiment, the UV source is turned on for 1 ms (200 pulses of 5 μs) to deposit a few electrons on the He surface. No signal appears even though the clocking sequence is running. This is due to the fact that our small gates form only a tiny fraction of the total channel area. The generated electrons will thus mostly find the large gates if they are emitted in the right direction or the walls of the cell (kept at 0 V as well) if they are not. The large gates therefore act as an electron collector and reservoir. When gate D is lowered to 0 V, a signal immediately appears as electrons enter the channels and are transported by the clocking voltage sequence. It is worth mentioning that we expect all the channels to carry an equal amount of charge since they are all connected together at the edge of the large gate, P1. The clocking signals we obtain in this manner are extremely stable (tens of minutes without a measurable decay). Bringing gate 2 to –2 V interrupts the signal by preventing the electrons from reaching or leaving the detection gate, 1. That signal is restored when gate 2 is brought back to 0 V thus showing the repeatability and stability of this system.

The number of electrons in the channels was reduced by lowering the depth of their potential wells (the top plane voltage was taken from –2 V to -0.4 V after charging the helium surface) and by closing gate D (bringing it to –2 V) so that electrons above the small gates could not be replenished from the reservoir if they ever escaped. The signal from gate 1 remained stable with an amplitude $V_s$=9 µV referred to the gate (using the measured gain of the amplifiers). The calculated capacitance of gate 1 and the input capacitance of the HEMT (NEC 3210S01) amount to about 1 pF, and thus the estimated number of electrons detected is $n=CV_s/e$ =56, or about one electron per channel.

The approach to measuring the charge transfer efficiency follows directly from ref. 10. One starts with a conventional 3-phase CCD clock sequence to move electrons back and forth between gates 1 and 7 as described earlier. However, after electrons are supposed to have left gate 7 on each cycle, gate 6 is made quite negative (-2.25V) so that any electrons remaining above gate 7 are trapped there. Then gate 7 is also made negative (-2.0V), expelling any trapped electrons from the system towards the top plane and the zinc-coated sapphire (made positive after the photoemission). Gate D was also held at -2V so that electrons could not enter from the reservoir to replenish those which did not transfer and were "kicked out" by this voltage sequence. Gate 2 was at 0 V to allow the electrons to reach the detection gate. Figure 3 shows the data obtained with this "kicking sequence." Points are spaced by 500,000 cycles (each cycle being a round trip from gate 1 to 7 and back) since we were taking one measurement every five seconds at 100 kHz. The solid curve is an exponential decay fit to the data and the extracted decay constant $\lambda$ gives $CTE=e^{\lambda}=0.99999992 \pm 6 \times 10^{-8}$. The same measurement repeated at 50 kHz returns a similar value, $CTE=0.99999989 \pm 5 \times 10^{-8}$. The 100 kHz frequency corresponds to round trips of the electrons, including pauses. To compare to a conventionally-defined CCD clocking frequency, the pauses must be removed and it must be taken into account that in one roundtrip the electrons traverse 4 "pixels" (3 gates form a pixel in a 3-phase CCD) during each cycle. Therefore, the clock (pixel) rate in these experiments is 800 kHz.

The charge transfer efficiency that we obtained is extremely high and noticeably higher than the 0.9990 figure extracted at 7 Hz in the previous measurements for electrons on millimeter-thick helium above macroscopic (3mm) gates.[10] The reason for the improved efficiency in the present experiments is twofold. First, the size of the gates is now of the same order of magnitude as the thickness of helium in the channel, allowing fringing fields to assist (slightly) the electron transport as in buried-channel Si CCDs. Second, the distance that electrons need to diffuse is now much shorter (microns instead of millimeters). All in all, electron transfer from one gate to another is much faster, and fewer electrons are left behind on gate 7 to be expelled.

If we solve the one-dimensional diffusion equation[14] for gate 7 using the diffusivity,

$D=17$ cm$^2$/s, that was obtained from the experiments on millimeter-thick helium, we find that the calculated CTE has ten nines after the decimal point. It is possible that very shallow potential traps are present which limit the measured CTE. Such traps could be explained by nonuniformities in the electrostatic field in the channels as might be caused by charged areas on the side insulator, for example. However, the diffusion calculation agrees with the experimental results if the diffusivity of the electrons is only about a factor of two less than we assumed above. Thus the presence of shallow traps is not certain.

At lower temperatures trapping due to nonuniformities could become more severe. However, fringing fields can extract charges from shallow traps. Analytical approximations for the fringing fields in CCD's have been developed and can be calculated by following Bakker.[15] Our present devices have gates which are 3-4 times wider than the depth of the channels, and thus the fringing fields are weak. For narrower gates the fringing fields are increased and negligible trapping is expected, even at much lower temperatures.

In summary, we have measured the charge transfer efficiency (CTE) of small numbers of electrons floating on a liquid helium CCD to be 0.99999992 at 800 kHz for 10 μm-wide gates. This figure is, to our knowledge, the highest CTE ever measured for a CCD made in any physical system (Si, GaAs or others) and is found to be sufficient for reliable quantum computing. Another way to look at the efficiency is that an electron fails to transfer from one gate to the next about once every 350 m for a structure with 10 μm gates. This CTE is even more impressive given the fact that on average the helium channel CCD has of the order of one electron per pixel rather than the thousands of electrons typical of high-CTE Si CCD's[16] . Very shallow traps may be limiting the current transfer efficiency. These traps might arise from imperfections in the lithography leading to modulation of the channel width, or local dielectric discontinuities leading to nonuniformities in the electric fields. The CTE of 0.99999992 obtained for our basic, non-optimized structure can likely be improved. Narrower gates, for example, will increase the fringing fields at the surface of the shallow helium channels, liberating electrons from any shallow traps and accelerating transport along the channels. The ability to clock few electrons reliably for long distances is necessary to evince scalability in quantum computing using electrons on helium. This work has demonstrated this capability.

This work was supported in part by the NSF under grant CCF-0323472, and by the ARO and DTO under contract W911NF-04-1-0398.

FIG. 1. (a) Pictures of the metal/polymer/metal structure seen from the top and (b) diagram of its vertical cross-section. Sixty parallel channels are filled with helium and the electrons at their surface are controlled by the underlying gates (eight small gates, numbered from 1 to 7, a door gate, labeled D and three large plates, labeled P1 to P3).

FIG. 2. Electrical connections to the sample. The large plates (P1 to P3) are used as an electron reservoir and to verify the presence of electrons. Gates 3 to 7 share a common dc potential but each has a different ac signal. Gates 2 and D are used as "doors" that prevent or allow electrons to move past them. Gate 1 is connected to a HEMT pre-amplifier and is used as the measurement gate.

FIG. 3. Signal from the electrons measured on a lock-in amplifier as a function of time (triangles) when the voltage sequence kicks out the electrons that fail to transfer out of gate 7 on each cycle. These points are fit to an exponential decay (solid line) assuming a CTE of 0.99999992.

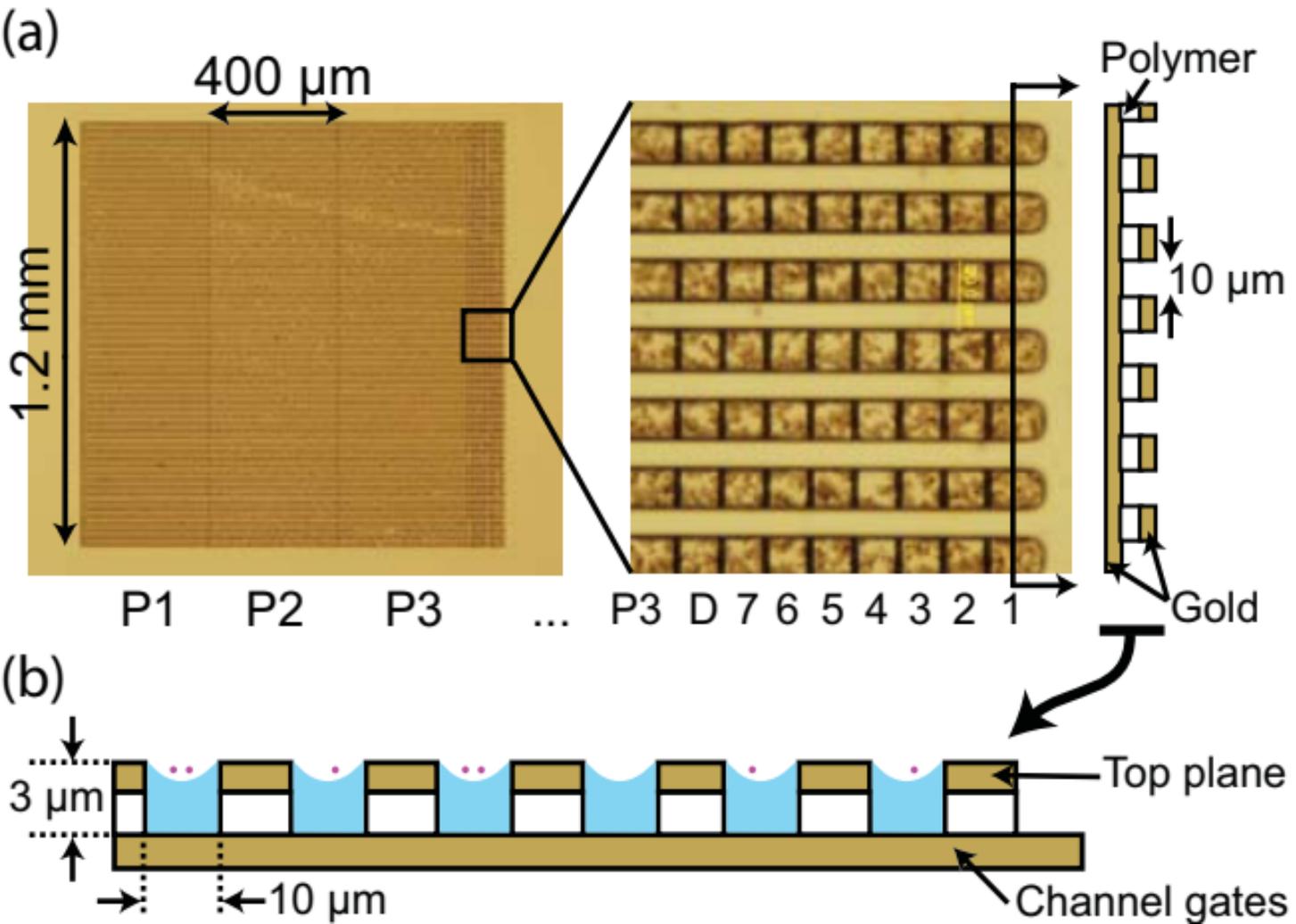

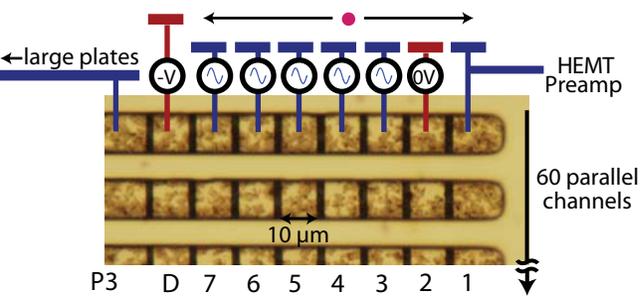

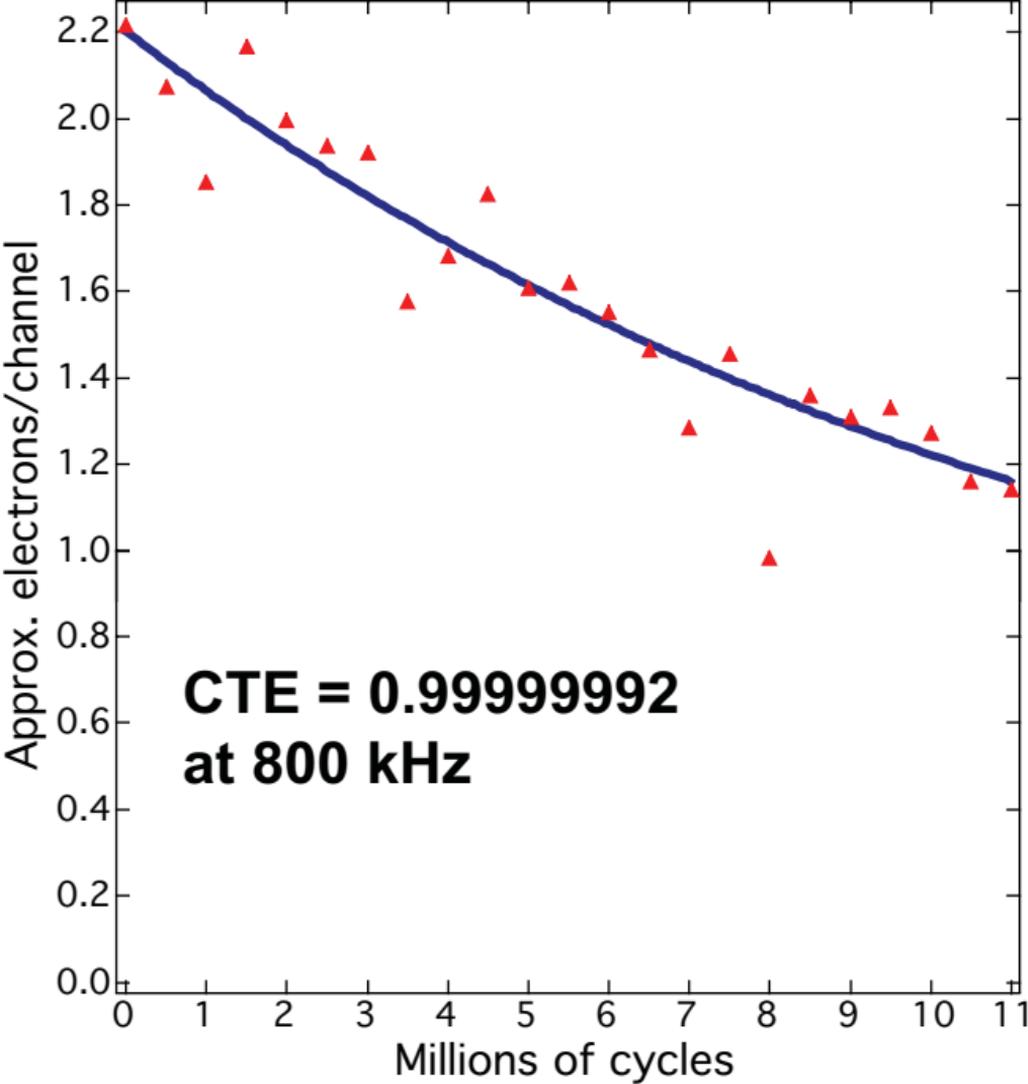